\documentclass[iop]{emulateapj}
\usepackage{graphicx}
\usepackage{wasysym}

\shorttitle{DRAFT: Red Runaways}
\shortauthors{Vickers et al.}

\begin{document}

\title{Red Runaways: Hypervelocity Stars, Hills Ejecta and Other Outliers in the F-to-M Star Regime}

\author{John J. Vickers\altaffilmark{1,2,$\dagger$}, Martin C. Smith\altaffilmark{2}, Eva K. Grebel\altaffilmark{1}}

\altaffiltext{1}{Astronomisches Rechen-Institut, Zentrum f\"{u}r Astronomie der Universit\"{a}t Heidelberg, M\"{o}nchhofstr. 12-14, 69120 Heidelberg, Germany}
\altaffiltext{2}{Key Laboratory for Research in Galaxies and Cosmology, Shanghai Astronomical Observatory, Chinese Academy of Sciences, 80 Nandan Road, Shanghai 200030, China}
\altaffiltext{$\dagger$}{johnjvickers@shao.ac.cn}

%
%
%Abstract
\begin{abstract}

In this paper we analyze a sample of metal-rich ($>$-0.8 dex) main sequence stars in the extended solar neighborhood, investigating kinematic outliers from the background population. The data, which are taken from the Sloan Digital Sky Survey, are kinematically profiled as a function of distance from the Galactic plane using full six dimensional phase space information. Each star is examined in the context of these kinematic profiles and likelihoods are assigned to quantify whether a star matches the underlying profile. Since some of these stars are likely to have been ejected from the disc, we trace back their orbits in order to determine potential ejection radii. We find that objects with low probability (i.e. `outliers') are typically more metal poor, faster and, most importantly, have a tendency to originate from the inner Galaxy compared to the underlying population.

We also compose a sample of stars with velocities exceeding the local escape velocity. Although we do not discount that our sample could be contaminated by objects with spurious proper motions, a number of stars appear to have been ejected from the disc with exceptionally high velocities. Some of these are consistent with being ejected from the spiral arms and hence are a rich resource for further study. Finally we look at objects whose orbits are consistent with them being ejected at high speeds from the Galactic center. Of these objects we find that one, J135855.65+552538.19, is inconsistent with halo, bulge and disk kinematics and could plausibly have been ejected from the Galactic nucleus via a Hills mechanism.

\end{abstract}

%
%
%Introduction
\section{Introduction}
The term `runaway' has traditionally been used to describe B type stars which are found in the Galactic halo. Since B type stars require gas rich, high density environments to form, their presence far from the Galactic disk is puzzling. Adding to the confusion is the observation that these stars may have intrinsically large peculiar velocities, which is difficult to explain if one assumes that these stars did in fact form in the Galactic disk. Several theories have been posited to explain how these stars may have been ejected from the populations they were born into.

The first theory is the so-called \emph{Binary Ejection Mechanism}. Proposed by \citet{bla1961}, this theory postulates that these runaway stars originated in binary systems and were `kicked' out of their systems by the death of their companions. If the companion goes supernova, its gravitational attraction will be greatly lessened by the shedding of mass, which can unbind the runaway and send it into the field. Additional effects may come from asymmetric explosions which are known to impart large kicks (sometimes in excess of of 1000 km s$^{-1}$) to the neutron star remnant \citep{sch2006}. This mechanism usually yields a runaway with a velocity similar to that of its pre-dissociated orbital velocity.

The second theory, the \emph{Dynamical Ejection Mechanism}, was suggested by \citet{pov1967}. In the case of binary systems interacting with another very massive star, one member of the binary may be captured by the interloping star, while the other could be ejected with high velocity (several hundred km s$^{-1}$ \citealt{gva2009big}). Similar effects could occur in interacting binaries \citep{leo1990}, or dynamically unstable tertiary systems. In these interactions, the kick imparted is usually close to the orbital speed of the binary components, but may be as large as the escape velocity from the surface of the most massive interacting object \citep{gva2009small}. This process is expected to dominate in crowded regions such as star clusters. In the low velocity regime, this mechanism is responsible for `stellar evaporation' and leads to the observed mass segregation in clusters.

A third ejection process, involving interaction of a binary system with a black hole, operates in much the same way as the dynamical ejection mechanism (in the case of a binary colliding with a massive star) but is capable of imparting much larger kicks due to the extreme mass of the black hole. This method of ejection is known as the \emph{Hills Mechanism} \citep{hil1988}. Stars ejected in such a manner were observationally confirmed first by \citet{bro2005} and later shown to have non-isotropic distributions \citep{bro2012}. This anisotropy is thought to be an effect of the accretion patterns of the supermassive black hole in the center-most regions of the Galaxy.

Up until very recently, studies of these ejected objects have been limited to high mass early type stars. This is because it is much easier to find high probability candidates for follow up. Looking at faint magnitudes and high latitudes, one composes a sample of short lifetime objects based on their colors; the faintness ensures a great distance for these intrinsically bright objects, and the high latitude ensures a Galactic halo location. Follow up spectroscopy is then needed to confirm the spectral type and to find the velocity and distance with high precision.

Now however, with large spectroscopic surveys such as the Sloan Digital Sky Survey (SDSS; \citealt{yor2000}), the RAdial Velocity Experiment (RAVE; \citealt{ste2006}) and the Large sky Area Multi-Object fiber Spectroscopic Telescope (LAMOST; \citealt{cui2012}), tremendous amounts of spectra are being collected over vast areas of the sky. When combined with long baseline astrometric measurements from the United States Naval Observatory (USNO; \citealt{mon2003}) catalog, proper motions may be calculated. All that is left is to estimate distances to objects using photometric distance estimations (such as isochrone fitting) and six dimensional phase space information for hundreds of thousands of objects can be estimated.

This allows for sophisticated data-mining efforts to find chancely observed outlier stars; such endeavors have already been undertaken by \citet{pal2013} who report some of the first G and K hypervelocity star candidates. We perform a search for escape-velocity stars using different methods and rediscover some of their candidates as well as report new ones.

This study constitutes one of the largest statistical studies of kinematically outlying late-type main sequence objects, enabled by exploiting modern million-item spectroscopic surveys and using kinematics to identify runaway populations. Prior studies of runaway stars have, by and large, been positionally based, and not kinematically based (candidates are selected via photometry only and then analyzed kinematically with follow up).

It is only recently that studies have begun to investigate low mass stars which may have been ejected from their neighborhoods of birth (see for example \citealt{pal2013}, \citealt{zie2015} and \citealt{zho2014}). Using a previously unexploited sample of red stars (typically F-to-M type), we assign them likelihoods based on their phase space coordinates and use these likelihoods to characterize them as a function of their probability of being runaways.

Owing to contradictory usage of terms in the literature, we would now like to declare the nomenclature used for the rest of this paper. When we use the term \emph{hypervelocity}, we mean that the star is traveling fast enough to escape the potential of the Milky Way. The term \emph{Hills star} will be used to denote a star which may have interacted with the central supermassive black hole. We will refer to stars which have kinematics similar with their neighbors as \emph{natural} and stars with unlikely kinematics will be referred to as \emph{outliers}.

In Section 2.1 we outline the data used in this study and characterize the reliability of the proper motion measurements. In Section 2.2 we explain the methods used to calculate object phase space information and orbital parameters. Section 3.1 explains the methodology we use to fit phase-space profiles to the data; and Section 3.2 details the usage of these fits to generate a likelihood value for every star based on its six dimensional position. Section 4.1 analyzes the relationship between this assigned likelihood and the characteristics of the stars: in Sections 4.2 and 4.3 we present possible hypervelocity stars and possible Hills stars. We conclude in Section 5.

\begin{table}
\begin{center}
\caption{Data Quality Cuts}
\begin{tabular}{cccc}
\tableline\tableline
& Lower Bound & Upper Bound \\
\tableline
u & 12.0 & 22.0 \\
g & 14.1 & 22.2 \\
r & 14.1 & 22.2 \\
i & 13.8 & 21.3 \\
z & 12.3 & 20.5 \\
($g$-$i$)$_{0}$ & 0.3 & 4.0 \\
\tableline
SSPP SNR & 10. & - \\
Fe/H & -0.8 (dex) & - \\
Fe/H Error & - & 0.14 (dex) \\
log(g) & 3.0 (dex) & - \\
log(g) Error & - & 0.32 (dex) \\
Radial Velocity Error & - & 3920.92 (km s$^{-1}$) \\
\tableline
Proper Motion R.A. Error & - & 11.15 (mas yr$^{-1}$) \\
Proper Motion Dec. Error & - & 11.15 (mas yr$^{-1}$) \\
\tableline
\label{tab:data_cuts}
\end{tabular}
\tablecomments{Note that once all cuts are applied, the maximal radial velocity error is actually 53.98 km s$^{-1}$. The apparent magnitude limits are chosen by the 95\% limiting magnitudes of the SDSS survey from the SDSS website and the saturation limits of the camera from Table 4 of \citet{gun1998}.}
\end{center}
\end{table}

\section{Data}

\subsection{Pipeline Products}

For our object sample, we utilize SDSS Data Release 10 objects with high quality spectroscopic parameters. To date, the SDSS has collected over a million science quality spectra, of which over 500,000 are point sources, the majority of these being stellar objects. These objects -- with proper motions, radial velocities, distance estimates (not from the SDSS pipeline products, but rather a separate method outlined below) and high-precision astrometry -- possess full six dimensional phase space information. This large set of phase-space data forms the basis of our analysis.

The SDSS is a long-running survey (the first data release being more than a decade ago, see \citealt{sto2002}) which, as of the tenth data release \citep{ahn2014}, has imaged over a third of the sky in five photometric bands ($u$, $g$, $r$, $i$, $z$: for information on the survey strategy, see \citealt{yor2000}; for information on the filters and imager, see \citealt{fuk1996} and \citealt{gun1998}, respectively). The SDSS 2.5 meter telescope, situated at Apache Point Observatory in New Mexico, has proved illuminating not only in its primary mission of extragalactic exploration, but also in elucidating the mysteries of our own Milky Way and its complex formation history (see for example \citealt{yan2000}, \citealt{new2002}, \citealt{bel2006} and references therein).

The SDSS telescope is also outfitted with twin multifiber spectrographs which can take up to 640 spectral readings simultaneously on 3" fibers. The spectrographs operate over the visual range (3900\AA\ to 9000\AA) at a moderate resolution (R$\sim$1850 - 2200). The Sloan Extension for Galactic Understanding and Exploration (SEGUE) project papers \citep{yan2009} and SDSS websites\footnote{http://www.sdss3.org/} outline the basic information pertaining to the stellar spectroscopy.

The SDSS offers a selection of stellar atmospheric parameters in its Sloan Stellar Parameter Pipeline data product (SSPP; see \citealt{lee2008}). In this work we limit our sample to spectra with signal to noise ratios of 10 or higher -- this corresponds to atmospheric parameter estimate errors of about: $\Delta T_{eff} \sim$ 103.9 K, $\Delta log(g) \sim$ .282 dex and $\Delta [Fe/H] \sim$ 0.213 dex (for the pipeline adopted values; see Table 6 of \citealt{lee2008}). We also use the radial velocity outputs from the SSPP. Radial velocities are generally found via ELODIE template matching (the ELODIE spectroscopic library is a collection of high resolution spectra collected by the ELODIE spectrograph at the Observatoire de Haute-Provence 1.93 m telescope; \citealt{mou2004}) and have accuracies from 4 to 15 km s$^{-1}$ (depending on the quality of the spectra, see Table 2 of \citealt{yan2009}). As a quality control measure, we reject any object which is critically flagged by the pipeline (see Table 7 of \citealt{lee2008} for a list of pipeline flags), while cautionary flags are allowed to pass.

The SDSS data products offer an internal proper motion table where proper motions are obtained by comparing USNO-B astrometry and SDSS astrometry \citep{mun2004}. The proper motions are obtained over the long time baseline between earlier photographic plate surveys and the SDSS observations and use the SDSS galaxy sample as a stationary reference frame. This catalog of proper motions is 90\% complete to $g \sim$ 19.7 (Note that our cuts produce a sample of objects of which 98\% are brighter than this magnitude).

In this paper we wish to focus on disk-origin objects which have non-disk kinematics. To weed out the halo, we investigate only objects with SSPP metallicities greater than -0.8 dex. An object with such a high metallicity, which is found to be inconsistent in its kinematics with its neighboring high metallicity stars, is a candidate runaway star.

For our distance estimates, we use a main sequence color-magnitude relation described by \citet{ive2008}; for this relation to be valid, the objects investigated must fall in the main sequence color range (0.3 $<$ ($g$-$i$)$_{0} <$ 4.0; the subscript `0' denotes a color that has been extinction corrected using the maps of \citealt{sch1998}) and  and have dwarf-like surface gravities ($log(g) > 3.0$). In essence these cuts select luminosity class V objects from early F to late M type. As a final quality cut, we also cull all data which have error measurements on any parameter which is more than 3$\sigma$ greater than the average error measurement of the data for that parameter (i.e. we cut off the long tails of the error distributions). A summary of the selection criteria for our final sample is outlined in Table \ref{tab:data_cuts}.

We note that proper motion measurements are probably the least reliable portion of this analysis. We have devoted an Appendix to discussing this.

\begin{figure*}
\includegraphics[width=\textwidth]{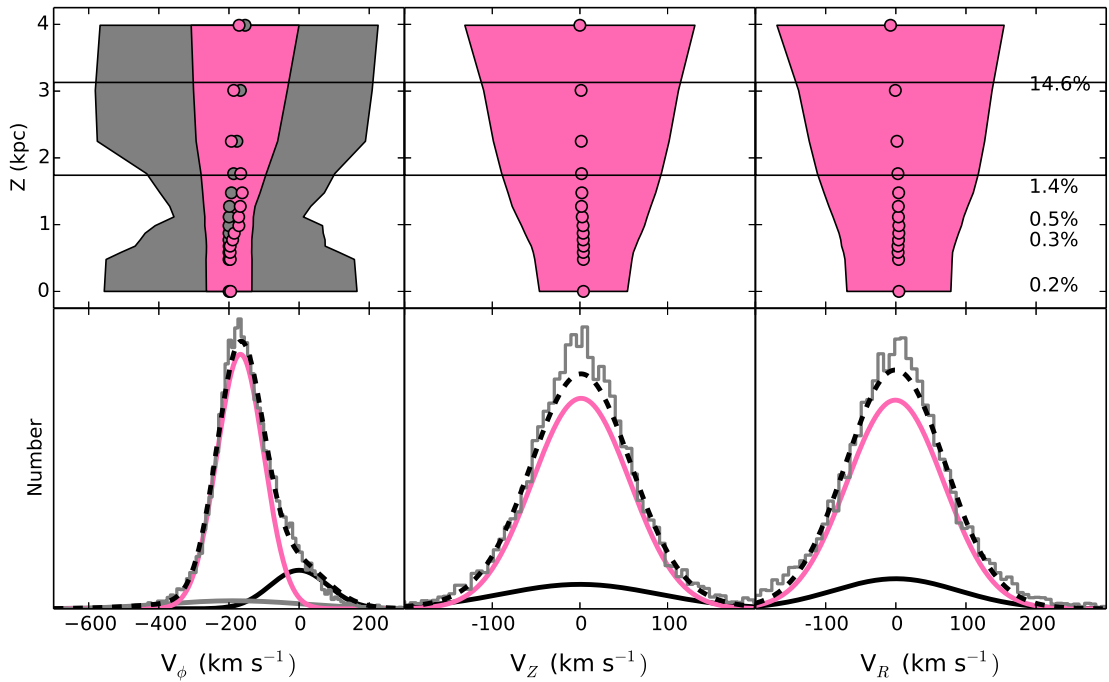}
\caption{\emph{Upper Panels}: The global fits to the velocity distributions of high metallicity stars at various distances from the Galactic plane. The shaded regions enclose the 2$\sigma$ ranges about the means (circles) for the fit Gaussians (pink). In the case of $v_{\phi}$, an additional Gaussian component (gray) is fit to the data to account for the heavy-tailed nature of that profile. \\
\emph{Lower Panels}: The fits for the data slice at Z$\sim$2.2 kpc (horizontal lines at 1.74 kpc and 3.13 kpc). The solid Gaussians correspond by color to the shaded regions described above (with an additional black Gaussian representing the halo). The data are shown with a gray histogram and the global fit is shown as a dashed line. \\
\emph{Annotations in the Top Right Panel:} Indicates the percentage of fits at various $Z$ values which is devoted to the halo. This percentage is determined by doubling the counter-rotating population.}
\label{fig:glob_fits}
\end{figure*}

\subsection{Calculated Distances, Velocities and Orbits}

Assigning distances to the objects is accomplished by a photometric estimate of the absolute magnitude of main sequence stars based on their ($g$-$i$)$_{0}$ colors. This estimate is given in parametric form in Appendix A of \citet{ive2008}. Their estimate is obtained: by fitting globular cluster photometry in combination with their distances as tabulated in the catalog of \citet{har1996} in the SDSS footprint for the bluest objects; with a combination of literature photometric distance estimate fits in the middle color regime; with Hipparcos M-dwarfs where possible more redward; and with ground based parallaxes for the reddest objects (see \citealt{ive2008} and \citealt{jur2008} for more details). This estimate accounts for metallicity by applying a polynomial correction to the estimated magnitude which is determined by comparing offsets of the estimated and true relations of a set of globular clusters as a function of the Harris Catalog metallicity estimates. These magnitudes are estimated to have a scatter of $\sim$ 0.1 to 0.2 -- this is accounted for later. We assign distances to our data by applying this photometric estimate and adding in the metallicity correction using SSPP spectroscopic metallicities.

For assigning Galactic Cartesian X, Y, and Z coordinates to objects, we assume that the sun has a position of -8.0 kpc and that the system is right handed (with X increasing toward the Galactic Center and Z increasing toward the Northern Galactic Pole).

Assigning velocities to the objects is straightforward. We initially assign Cartesian U,V,W velocities (= dX/dt, dY/dt, dZ/dt) to the objects, in a manner similar to \citet{joh1987}\footnote{see $http://idlastro.gsfc.nasa.gov/ftp/pro/astro/gal\_uvw.pro$ for example IDL code} using the prior outlined proper motions and radial velocities.

The cylindrical rotational ($v_{\phi}$), radial ($v_{R}$) and vertical velocities ($v_{Z}$) are then calculated in the standard fashion. The direction of $v_{\phi}$ is chosen to be right handed (i.e. the Solar rotation speed is -220 km s$^{-1}$).

In short, all systems used are right handed with Galactocentric origins.

We calculate the orbital paths for each star in the data set by integrating their positions and velocities through the potential presented in \citet{deh1998}. Of the 35 available potentials, we use 2b. This potential was chosen by rough comparison of the Galactic parameters in Tables 3 and 4 of \citet{deh1998} with Tables 2 and 3 of the more recent study by \citet{mcm2011}.

In this model, the disk is composed of 3 exponential density constituents: an interstellar medium component, and the thin and thick stellar disks. The bulge and halo density profiles are each described by spheroidal functions. Then the total gravitational potential is made to satisfy Poisson's equation and is brought into agreement with observational constraints such as the Milky Way's rotation curve and peak velocities of the interstellar medium. For full details, please refer to the paper of \citet{deh1998}; the code is made available as part of the NEMO Stellar Dynamics Toolbox\footnote{http://carma.astro.umd.edu/nemo/}.

The orbits are calculated for every star in our data set by reversing their U,V,W velocities and running them through the potential for 14 Gyr at a 1 Myr resolution. The choice of a 14 Gyr integration time is to ensure we find crossings where possible. This does not account for a changing Milky Way potential, but since the average lookback time to the previous crossing is $\sim$39 Myr, we do not consider this to be an issue. When a planar crossing is detected, a linear interpolation between the calculation steps before and after the crossing is used to find the exact coordinates and velocities of the crossing point.

\section{Fitting and Likelihoods}

\subsection{Fitting Kinematic Profiles to the Data}
To find outlying objects, we must first define the expected distribution of stars in phase space. To do this, we fit the kinematic properties of the objects ($v_{R}$, $v_{Z}$ and $v_{\phi}$) at different distances from the plane of the Galaxy to Gaussian mixture profiles. By interpolating between these fits, we can infer the expected velocity profile of an object at any distance from the plane.

Fitting is performed by cutting the data set into $Z$ slices with 20,000 members per slice (after the data quality checks, we have about 135,000 objects) and the fits are boxcar smoothed with a step size of 10,000 (so the data are cut into 14 slices each of which have 10,000 members in common with the previous slice, and 10,000 in common with the next slice). This dynamic binning offers a similar robustness of fit for each slice (excepting the most distant two slices), but suffers from a differential resolution of the fits, with the more distant bins growing wider and wider apart due to less complete sampling (notice that very close bins are also wider apart due to the bright limit of the SDSS survey). The median bin size is about 0.32 kpc. A mixture of Gaussians is then fit to the members of each slice in each cylindrical velocity component.

$v_{\phi}$ is the most complex profile to fit. The distribution of rotation speeds in the disk is known to be non-Gaussian, with a heavy tail extending to slower rotation velocities (due to asymmetric drift -- the effect where $v_{\phi}$ is inversely correlated to the random motion of a star; see Section 10.3.1 of Binney \& Merrifield 1998 pp. 624-629). This can be described using physically motivated models, such as the one presented in \citet{sch2012}, or  simply by fitting two Gaussians to the disk population in $v_{\phi}$ (which is the approach we take). As the $Z$ value of the fitting slice increases, a counter rotating population becomes increasingly evident. Despite the high metallicity cut, some halo population appears to be leaking into our sample. To rectify this, a halo population is added into the fit. The halo population's membership is defined to be twice that of the counter rotating population (the equivalent of assuming the halo population is non-rotating and that the entire counter rotating population has halo membership). Moreover, the halo population is assumed to have a dispersion invariant with respect to $Z$ and no net motion.

The $v_{R}$ and $v_{Z}$ distributions of the disk are much more symmetric, so they are each fit to single Gaussians. An additional halo component is also added to these fits, with the normalization of this component again being determined by the counter rotating population; again we assume no net motion and a constant dispersion of velocities at all heights from the plane.

The fitting is done via a maximum likelihood method using the Markov Chain Monte Carlo technique, carried out on each Z slice in turn. To begin with each velocity component in the slice is fit with the following probability density function:

\begin{equation}
P(v_{k}) = \sum_{j=1}^{N} A_{j} \frac{1}{\sigma_{j} \sqrt{2\pi}} e^{\frac{-(v_{k}-\mu_{j})^2}{2\sigma_{j}^{2}}},
\label{eqn:prob}
\end{equation}

where k corresponds to the velocity component (i.e. $v_{R}$, $v_{\phi}$, $v_{Z}$), N is the number of Gaussian components in the fit (i.e. N = 2 for $v_{R}$ and $v_{Z}$, N = 3 for $v_{\phi}$), and $A_{j}$ is the normalization of each Gaussian. The halo Gaussian is mostly fixed, with $\mu_{R}$ = $\mu_{\Phi}$ = $\mu_{Z}$ = 0 km s$^{-1}$ and $\sigma_{R}$ = 138.2 km s$^{-1}$, $\sigma_{\Phi}$ = 82.4 km s$^{-1}$ and $\sigma_{Z}$ = 89.3 km s$^{-1}$ \citep{smi2009}. The normalization of the halo is determined from the fit to $v_{\phi}$ (see above).

The parameters for the above fits are determined by carrying out a maximum likelihood calculation, with the likelihood for each velocity component k given by:

\begin{equation}
\mathcal{L} = \prod_{i=1}^{N_{star}} P(v_{k}^{i}),
\label{eqn:likelihood}
\end{equation}

where the product is carried out over the total number of stars in the slice (i.e. $N_{star}$ = 20,000).

\begin{figure}[h]
\includegraphics[width=\linewidth]{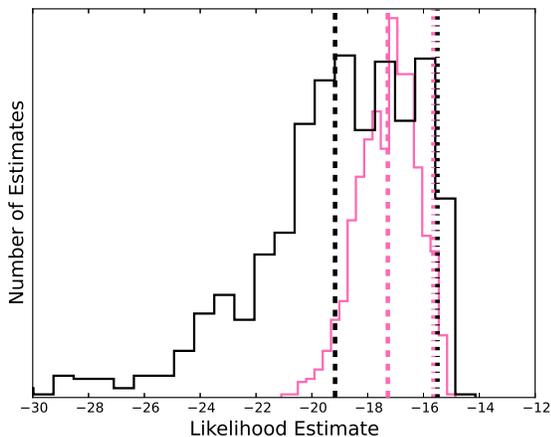}
\caption{Example of how we assign likelihoods to stars. In this figure we show two Monte Carlo expansions of the same object. The pink histogram represents the likelihood distribution of 1000 realizations of the star and the dashed and dotted pink lines represent the average likelihood and 95\% highest likelihood of these expansions (respectively). The black histogram, dashed and dotted lines are the same except the errors on the base star's measurements have been inflated threefold. We see that the object with large errors would have a much lower average likelihood than the object with small errors, despite being from the same underlying observations. However both expansions have very similar 95\% likelihoods. We use the 95\% likelihood instead of the average likelihood in order to ensure that our outlier stars are true and not classified as outliers merely because of large errors.}
\label{fig:likelihood_stretch}
\end{figure}
The global fits are depicted in Figure \ref{fig:glob_fits} (the means of the fitted Gaussians are plotted as points, and the standard deviations are depicted by filled areas as a function of distance from the plane).

It should be noted that these fits are not intended to provide any physical insights into the disk. These fits are just probability density functions of high metallicity stars and are used to find kinematic outliers within this high metallicity population. For more proper analyses of disk population kinematics, we direct the reader to \citet{smi2012}, \citet{sch2012}, and the review of \citet{rix2013}.

\subsection{Assigning Likelihoods to Stars Based on the Kinematic Profiles}

\begin{figure}[h]
\includegraphics[width=\linewidth]{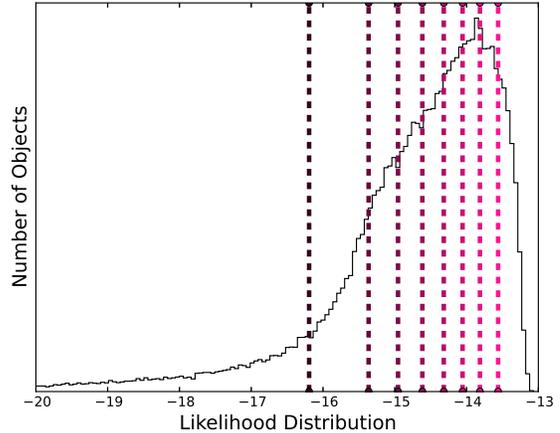}
\caption{The likelihood distribution for all objects in our sample. Objects with higher likelihoods exhibit disk like motions. Objects with low likelihoods have incongruous phase space values compared to the rest of the objects in their vicinity (for example they may be counter-rotating or have high W velocities). The vertical lines indicate divisions between 9 equally populated likelihood bins for the stars. These data are compared in Figures \ref{fig:likelihood_histograms} and \ref{fig:combined}, and the division colors here roughly correspond to the color gradients in those figures.}
\label{fig:likelihood_distribution}
\end{figure}

To find outliers, we use the Gaussian fits from the previous section as probability density functions. These probability density functions, which vary as a function of height from the plane, fit the overall distributions of our high metallicity stars. Outliers in these fits are likely also outliers of the underlying population.

\begin{figure*}
\includegraphics[width=\textwidth]{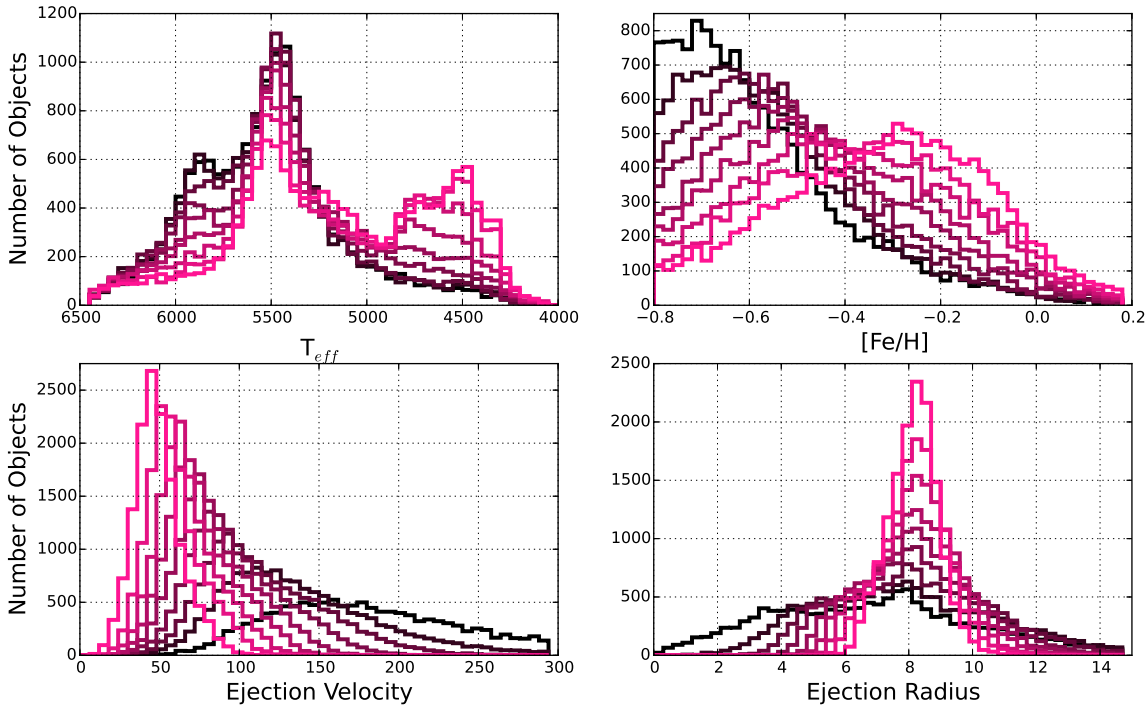}
\caption{Trends in fundamental properties (T$_{eff}$, [Fe/H]) and orbital crossing properties (crossing velocity and Galactic radius) of the stars as a function of their likelihood (see Figure \ref{fig:likelihood_distribution}). We see that outliers (black) are: hotter, more metal deficient, have faster crossing speeds, and more probably crossed last at smaller radii than the natural stars (pink).}

\label{fig:likelihood_histograms}
\end{figure*}

\begin{figure}
\includegraphics[width=\linewidth]{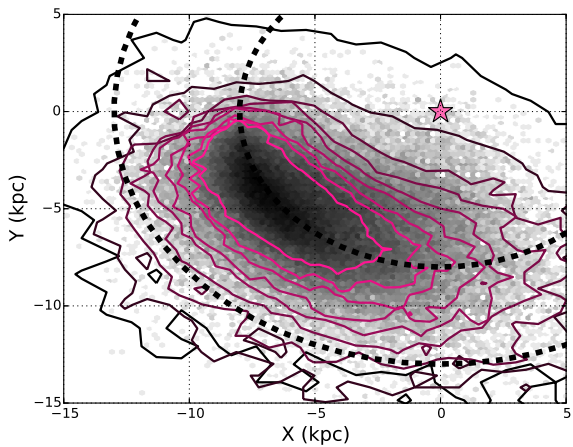}
\caption{The last point of Galactic plane crossing for our objects. The grayscale density is a logarithmically weighted hexbin of all of the objects' crossing points. The objects that we observe seem most densely concentrated on an orbit similar to that of the Sun. Notice that there is a surplus of objects observed coming from the Galactic center as opposed to traveling inward. The inner dashed line indicates an 8 kpc radius circle about the Galactic center (the star). The outer dashed line traces a circle with a 13 kpc radius, where the density of the disk begins to drop dramatically in early type stars according to \citet{sal2010}. The contours follow the color scheme of the prior two plots: natural stars are indicated in pink, and outliers are black. The contours show the 2$\sigma$ limits for the crossing areas of the objects. The natural stars follow the local rotation, while the outliers show a preference to originate from more Galactocentric regions.}
\label{fig:combined}
\end{figure}

In order to quantify the likelihood that the $i^{th}$ star is consistent with the underlying distribution, we calculate the product of the probabilities for each velocity component:

\begin{equation}
P_{i} = P(v_{R}^{i}) P(v_{\phi}^{i}) P(v_{Z}^{i}).
\label{eqn:tot_prob}
\end{equation}

Here the probabilities for each velocity component are calculated using Equation (1), taking the best-fit parameters determined for the relevant slice in z. Clearly velocity outliers, such as stars ejected from the disc, will have low values of P$_{i}$.

We now incorporate the errors on the observational measurements which have, until this point, been neglected. For each individual star we create a sample of 500 error-incorporated realizations by Monte Carlo sampling the relevant measurements (g, r, i, [Fe/H] for the distances; radial velocity, and proper motions for the kinematics). An additional 0.15 mag error is factored into the distance estimates to account for the uncertainty in the color-magnitude relation \citep{ive2008}. Each realization is assigned a likelihood using Equation (3) and in such a manner a distribution of likelihoods for each star is constructed.

One of the largest concerns of our analysis is the presence of large uncertainties in the observational measurements; for example, a star with a large uncertainty in its proper motion could be reported to have a very large spatial velocity, even though the large uncertainty means that it could be consistent with a more moderate velocity. By using this Monte Carlo approach to obtain a distribution for P$_{i}$, rather than an individual value, means that we can weed out stars with large uncertainties. This is important because the non-linearity of the velocity calculation means that the Monte-Carlo likelihood distributions could be very asymmetric, with large tails to low probabilities. This will have greatest impact on the stars in the high-velocity tails of the velocity distributions, which are the very stars we are most interested in.

To overcome this problem, for each star we adopt as its likelihood the 95\% highest likelihood from the Monte Carlo resamples. For objects with small uncertainties, this adopted likelihood will be close to the average likelihood of the distribution. For objects with large uncertainties, the adopted likelihood will be much higher than the average. This helps to clean our final sample of low likelihood stars, removing objects which have low likelihoods merely by virtue of their large uncertainties. This is illustrated in Figure \ref{fig:likelihood_stretch}.

Figure \ref{fig:likelihood_distribution} depicts the distribution of 95\% likelihoods for all of the stars we investigate.

\begin{figure*}
\includegraphics[width=\textwidth]{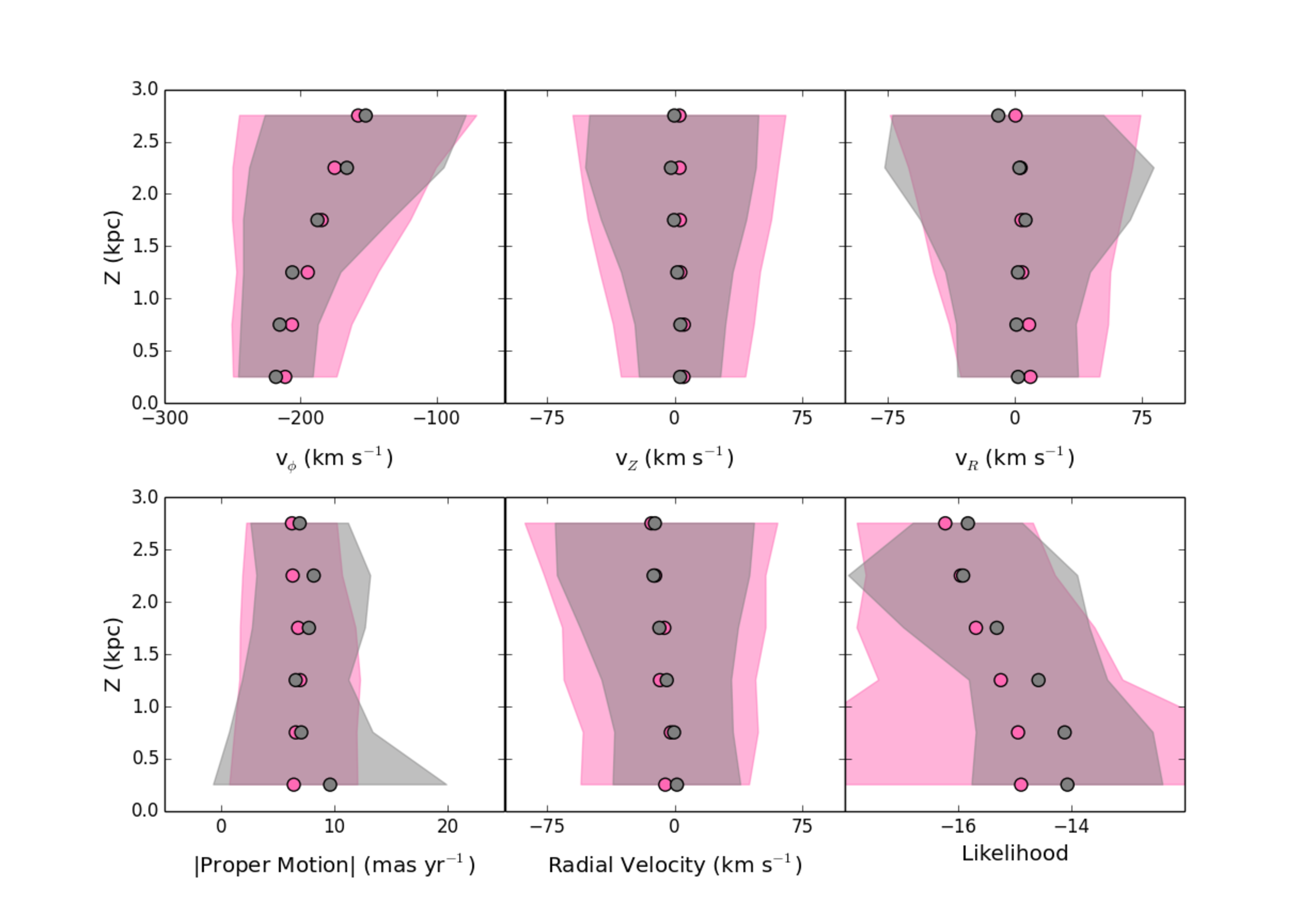}
\caption{\emph{Upper Panels}: Cylindrical velocities as a function of distance from the Galactic plane. The average cylindrical velocity (points) and dispersions (shaded areas) are shown for early type (T$_{eff}$ $>$ 5700 K) and late type (T$_{eff}$ $<$ 5000 K) stars (pink and gray, respectively). We see that the rotational velocities of the two populations are consistent, which implies that the distance estimates are unbiased. We also see that the kinematics of early type stars are usually more dispersed (especially in $v_{\phi}$ and $v_{Z}$) which would explain their generally lower likelihoods. \\
\emph{Lower Panels}: The measured proper motion magnitudes (left) and radial velocity magnitudes (middle). We see that the early type stars are more dispersed in radial velocities than the late type objects, while having similar proper motion measurements. Since the radial velocities are much more reliable than the proper motions, this implies that the early type stars are actually kinematically hotter than the late type stars. On the right we have the likelihood calculation as a function of $Z$. We see that the early type stars are less likely than the late type stars at all distances from the plane-- this is consistent with a dynamically hotter population.}
\label{fig:f_vs_k}
\end{figure*}

\section{Analysis}
\subsection{General Characteristics}

Having assigned likelihoods to all of the objects based on their phase space information, we now wish to investigate the attributes of the outliers in relation to the natural stars. To do so, we split the data into 9 equally populated partitions based on their calculated probability (see Figure \ref{fig:likelihood_distribution}).

We also investigate the crossing velocities and crossing radii of these objects. By integrating their orbits back in time, we can calculate approximately where they last crossed the Galactic plane and their velocity at this point. We refer to the radius of their last crossing point (with respect to the center of the Galaxy) as the `ejection radius' and to the velocity with which they crossed the plane as the `ejection velocity.' We correct the ejection velocity by accounting for the rotation curve of the Galaxy presented in \citet{cle1985} (so the ejection velocity indicates the crossing velocity with respect to the local standard of rest [LSR] at the crossing distance; note that we scaled the fit in that paper to have a solar position of 8 kpc instead of 8.5 kpc, to maintain the internal consistency of this paper).

Note that we assume here that the last crossing time was the point where the object was ejected, if it was ejected. The outliers are our main candidates for objects being involved in possible ejection events and form the basis for our analysis of ejection mechanisms acting on main sequence stars. There is no clean cut between outliers and natural stars-- instead it is a smooth transition and only general properties can be investigated.

In Figure \ref{fig:likelihood_histograms}, we histogram the characteristics (ejection radius, ejection velocity, metallicity, and temperature) of all of the objects we investigated in nine equally populated, color coded bins (the cuts again being shown in Figure \ref{fig:likelihood_distribution}). From these figures, we note four trends:

1) The natural stars are generally of later stellar types than the outliers. The population of G dwarfs is primarily orbiting more or less normally, while the population of early F dwarfs is more anomalous in its kinematics.

2) In general, the higher the metallicity, the more normal the rotation.

3) As expected, objects exiting the disk with large velocities compared to the LSR are less likely than objects rotating along with the LSR.

4) The natural stars, probably in orbits very similar to that of the Sun, have last crossed the disk at radii close to that of the Sun's orbit. The outliers have more varied crossing positions, but tend to come from closer to the Galactic center.

So, the natural objects are low temperature, metal rich, slow, and on orbits coincident with that of the Sun. The outliers are of earlier stellar types, metal poor, fast, and have a greater variance in crossing positions (with a tendency to be traveling outward from the more central regions). While the metal poor end of our data may suffer some halo contamination, the anisotropic velocities (i.e. the skewed distribution of ejection radius) suggest that they are predominantly disk or bulge stars being expelled from the central regions.

In Figure \ref{fig:combined} we show the relationship between the last crossing position of objects and their likelihoods. The grayscale density is a logarithmically weighted density plot of the last crossing positions of all the stars in our sample. The colored contours show the 2$\sigma$ areas of crossing for the objects as a function of their likelihood. The color scheme of these contours is the same as in Figures \ref{fig:likelihood_distribution} and \ref{fig:likelihood_histograms} (natural stars are pink, outliers are black). From this figure we can see how the outliers tend to be traveling outward from the central regions of the Galaxy; while natural stars tend to be traveling along the local rotation vector.

Here we would like to reiterate that, while the outliers exhibit many halo-esque characteristics such as low metallicity and varied crossing positions, and while it can be difficult to individually distinguish thick disk stars from halo stars in the metal poor end of the distribution we investigate here, we do not expect the halo to be a major contributor to this portion of the analysis. We refer to Figure \ref{fig:glob_fits} which is annotated with approximate halo contamination levels as a function of distance from the plane in our data set, as determined by the number of counter-rotating stars, see Section 3.1 for details. If the methodology leading to that estimate is to be believed, then the halo fraction does not exceed 27.5\% at any height from the plane in our sample; and the total halo contamination, estimated by averaging the contaminations in each z bin, is 4\%.

\subsubsection{A Case Against a Temperature Based Systematic Bias}
The trend for likelihoods to differ with temperature could be an indication of an underlying bias in some of our calculations. More specifically, there is a worry that our distance estimates are systematically too high or too low for certain spectral types. To investigate this, in Figure \ref{fig:f_vs_k} we plot calculated cylindrical velocities, proper motions, radial velocities and calculated likelihoods as a function of distance from the Galactic plane. The data are split into two groups: stars with temperatures below 5000 K, and stars with temperatures above 5700 K.

We find that at all distances from the plane, the rotational velocity, $v_{\phi}$, is consistent between the two populations with the early type stars having slightly larger spreads in their velocities. This implies that neither stellar group is being over or under estimated in distance with respect to the other. Further, we find that at all distances from the plane, the proper motions are similar in magnitude and dispersion; but the radial velocity estimates for the early type stars are more dispersed. This means that the calculated dispersions in the velocities are an effect of the precise radial velocity measurements and not due to some problem with the proper motion measurements. This radial velocity dispersion difference is manifest in the $v_{Z}$ measurements, where the early type stars are more dispersed than the late type stars (considering that our survey area consists of two cones directed out of the plane).

It is unexpected that the earlier type stars would have intrinsically larger velocity spreads than later types. Looking at the radial velocity error distributions of the early type and late type stars, we notice that the early type stars have generally larger errors than late type stars (2.4 km s$^{-1}$ as opposed to 1.4 km s$^{-1}$). This is perturbing as it suggests that the trends in Figures \ref{fig:likelihood_histograms} and \ref{fig:combined} may merely be a side effect of erroneous measurements. We investigate this further by performing Monte Carlo error expansions on several randomly selected stars, varying the magnitude of the radial velocity uncertainties. It turns out that larger radial velocity errors end up increasing a star's 95\% likelihood; so the lower likelihood of the earlier type stars is most likely an intrinsic property rather than a side effect of the measurement errors. We discuss possible causes of this trend in the conclusions.

\subsection{Hypervelocity Stars}

We identify hypervelocity stars as stars whose current kinetic energy is greater than the gravitational potential (described in Section 2.2) at their position. This selection allows for a changing selection threshold as a function of Galactic position and thus is more inclusive and correct than a constant velocity cut.

We construct a catalog of objects which satisfy this criterion and present a subset of them in Table \ref{tab:hv_candidates}. This subset is the group of escape velocity objects which also meet the following criteria:

1) their SDSS proper motions differ from PPMXL \citep{roe2010} proper motions by less than 12 mas yr$^{-1}$,

2) their absolute SDSS proper motions are less than 30 mas yr$^{-1}$,

3) their nearest observational (in projection) neighbor is more than 10 arcseconds away,

4) their SDSS proper motion errors are less than 5 mas yr$^{-1}$.

This choice of criteria is explained in the Appendix. Seven stars pass all of these quality cuts and 35 fail just one of these tests. The full sample of objects collected is presented online along with how many flags each object raised. In the online table, we rediscover seven of the thirteen candidates of \citet{pal2013}; five of their candidates are too metal poor to be included in our study, and one has large proper motion errors which exclude it from our data set (see Table \ref{tab:data_cuts}).

\begin{table*}
\begin{center}
\caption{Hypervelocity Candidates}
\begin{tabular}{ccccccccc}
\tableline\tableline
IAU Name & $\Delta$ P.M. & Nearest Neighbor & Total P.M & $\sigma$ P.M. & [Fe/H] & V$_{Total}$ & V$_{esc}$ & Infalling \\
 & mas yr$^{-1}$ & arcseconds & mas yr$^{-1}$ & mas yr$^{-1}$ & dex & km s$^{-1}$ & km s$^{-1}$ & \\
\tableline
J082015.88+362223.26 & 9.05 & 24.2 & 24.27 & 4.7 & -0.73 & 693.64 $\pm$ 104.32 & 523.83 & X \\
J160707.30+372350.97 & 9.13 & 19.8 & 15.83 & 4.36 & -0.11 & 606.47 $\pm$ 121.33 & 548.79 & \\
J131238.82+393312.13 & 8.57 & 11.13 & 22.96 & 4.75 & -0.63 & 607.16 $\pm$ 133.24 & 531.07 & \\
J175010.68+262448.38 & 4.72 & 13.32 & 20.9 & 4.47 & -0.52 & 694.04 $\pm$ 121.97 & 577.02 & \\
J075516.37+662909.15 & 9.25 & 16.57 & 8.95 & 4.94 & -0.72 & 620.95 $\pm$ 146.57 & 506.21 & X \\
J024605.10+312254.29 & 4.04 & 16.22 & 11.0 & 4.94 & -0.27 & 654.03 $\pm$ 137.45 & 518.32 & X \\
J093103.04+134728.81 & 5.63 & 12.95 & 9.1 & 4.55 & -0.51 & 577.78 $\pm$ 142.13 & 509.41 & \\
\tableline
\label{tab:hv_candidates}
\end{tabular}
\tablecomments{Objects whose kinetic energy is greater than the gravitational potential at their given position. This table is truncated to retain only the candidates that pass our strict proper motion quality cuts -- the full table (including objects which fail the criteria presented in the text) is available as an online supplement. Note that three of our candidates appear to be of extragalactic origin.}
\end{center}
\end{table*}

\begin{figure}
\includegraphics[width=\linewidth]{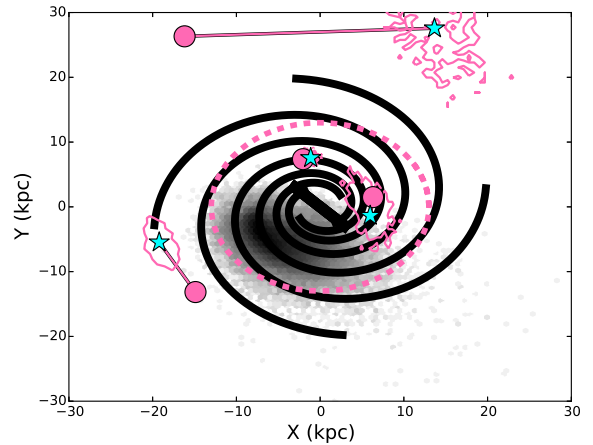}
\caption{The positions of last intersection with the Galactic plane for our hypervelocity candidates which intersect the Galactic plane. The grayscale density indicates the locations of last crossing for our entire data set. The current positions of the central bar and the spiral arms of the Milky Way are shown in black with a thickness about equal to 800 pc; \citet{val2014} determine the arms to have $\sim$800 pc diameters. The pink circles are the intersection points of the hypervelocity candidates; they are linked via pink vectors to their intersection points with respect to the spiral arm structure at the time of ejection (cyan stars). The contours are the one sigma distributions of the intersection positions. The dashed pink circle indicates the radius at which early type stars begin to diminish drastically in density as noted by \citet{sal2010}. Two of these objects are consistent with having been ejected from the spiral arm features of the Milky Way -- one intersects out of the expected radius and between spiral arms, and one intersects far outside of the expected extent of the disk.}
\label{fig:escapers}
\end{figure}

As an alternate test of the validity of our candidates, we refer to some fascinating work done by \citet{sil2013}. They effectively traced the positions of the Milky Way's spiral arm structures by tracing B type hypervelocity star orbits backwards in time. This takes advantage of the fact that both the binary and dynamical ejection mechanisms will occur more frequently within the crowded spiral arm structures. In Figure \ref{fig:escapers} we present a similar analysis with our sample of objects.

Figure \ref{fig:escapers} is a map of the bar (shown as a 3.1 kpc long feature inclined at 32$^{\circ}$ from the sun, as in \citealt{val2008}) and spiral arm structure of the Milky Way. The logarithmic spiral arms are described by the equations:

\begin{equation}
X = -Rsin(\theta) ,\; Y = Rcos(\theta) ,\; R = R_{0}e^{k\psi},
\label{eqn:spiral_1}
\end{equation}

\begin{equation}
k = tan(p) ,\; \psi = \theta - [0^{\circ}, -90^{\circ}, -180^{\circ}, -270^{\circ}] + \phi,
\label{eqn:spiral_2}
\end{equation}

where $p$ is the pitch angle of the arms, $\phi$ is the phase angle offset of the arms and $R_{0}$ is a characteristic radius. The four possible angles for the value of $\psi$ correspond to the four spiral arms of the Milky Way. In our case: $p = 12.8^{\circ}$, $\phi = -53.1^{\circ}$ and $R_{0} = 2.1$ kpc. These values were originally derived by \citet{val2008} for a Solar radius of 7.6 kpc, but \citet{sil2013} corrected them for a Solar radius of 8 kpc and presented them in their paper.

However, this is not the full picture. The spiral arm structure is rotating in time as well with a characteristic pattern speed $\Omega = 20.3$ km s$^{-1}$ kpc$^{-1}$. So when we see an object which has traveled for some time, its intersection point will be shifted with respect to Galactic structure. To rectify this, we rotate all the intersection points forward by the pattern speed multiplied by the time in travel. The intersection points with respect to the spiral structure at the time of ejection are shown with the cyan stars and connected to their actual intersection points by the pink vectors.

Two of our objects appear consistent with possible spiral arm ejections (considering errors), one intersects the Galactic plane so far away that it is doubtful it interacted with any major Galactic structures, and one intersects relatively far out and between spiral arms. The average ejection velocity of the three Galactic objects, with the Galactic rotation considered, is 669.5 km s$^{-1}$. This is on the high end of the scale of predicted ejection velocities for B type stars from the dynamic and binary ejection mechanisms, but is not unheard of; see, for example, \citet{gva2009small}. However, since maximal achievable velocities are inversely correlated with the mass of the ejected star, these velocities are consistent with maximum ejection speeds obtainable by our lower mass stars in binary ejections (up to 1000 km s$^{-1}$, \citealt{tau2015}).

\begin{figure}
\includegraphics[width=\linewidth]{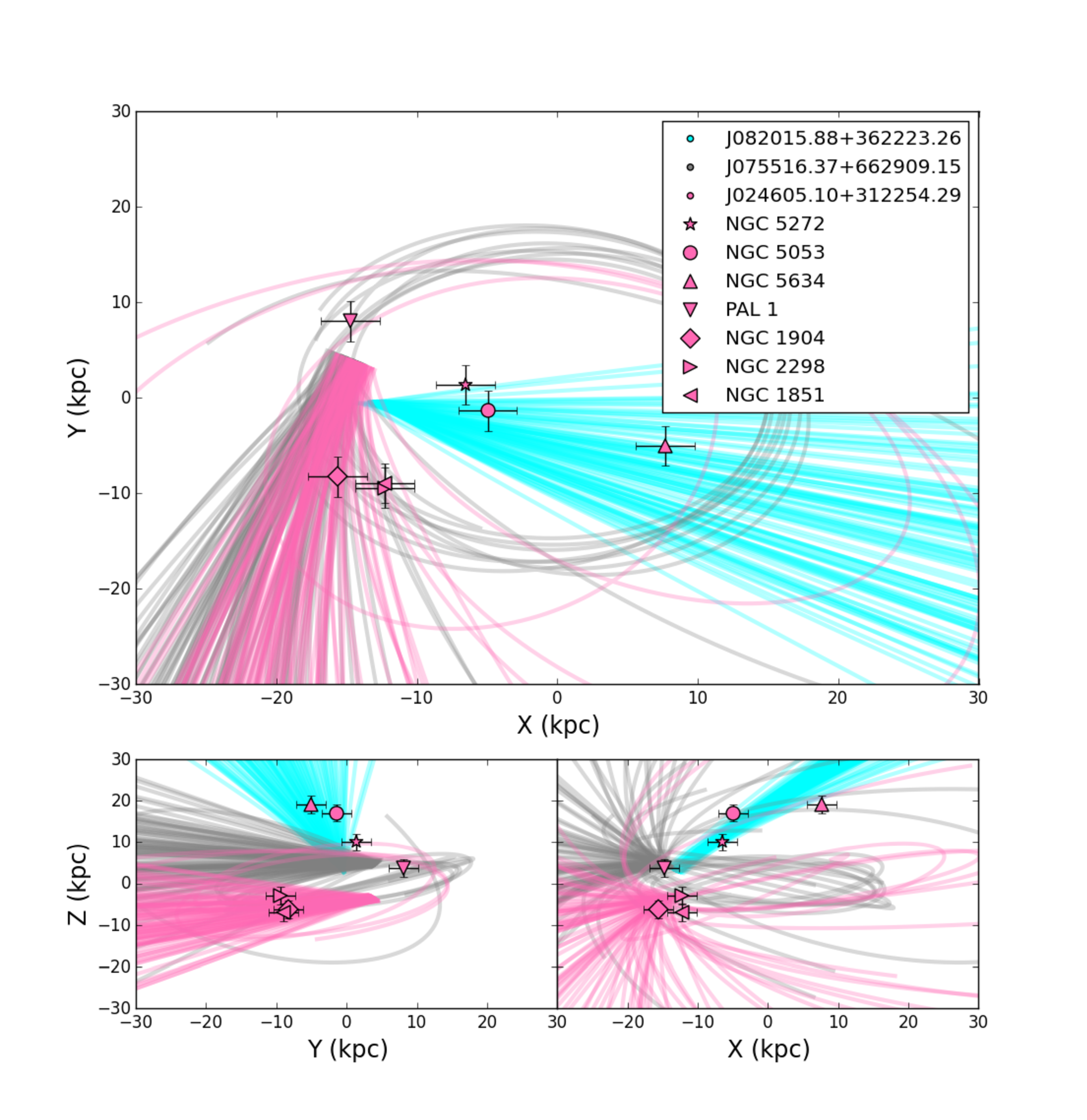}
\caption{Monte Carlo orbit expansions of three unbound, infalling, high metallicity stars. Also plotted are globular clusters which have positions within 5 kpc of the nominal orbits of the objects. The object J024605.10+312254.29 (pink) has an orbit coincident with the position of the globular cluster NGC 1904; however the star and the cluster differ in metallicity by 1.33 dex which is mysterious. The other objects are not strongly associated with any known globular clusters. The error bars indicate the average movement range of a globular cluster in the flight timescales considered here (17 Myr). }
\label{fig:infallers}
\end{figure}

We also wish to investigate the stars which did not cross the Galactic plane in their orbit integrations. These are perhaps the most mysterious stars because of the high metallicities of our sample; these high metallicities make it unlikely that these objects are merely halo contamination. Tracing their orbits back in time could provide insight into other systems capable of producing hypervelocity stars and constrain high velocity ejection mechanisms. In Figure \ref{fig:infallers} we show Monte Carlo orbit realizations for these stars along with nearby globular clusters from the catalog of \citet{har1996}.

While we do not have full phase velocity for all of these globular clusters, we can estimate errors on their spatial positions relative to the Monte Carlo orbit expansions. To do so we collect proper motions for a sample of globular clusters (five from \citet{din1}, ten from \citet{din2}, seven from \citet{din4}, six from \citet{din5}, nine from \citet{din6}, three from \citet{din7}, and 19 from the literature compilation of \citet{din3}) and find their total space velocity by comparison with radial velocities presented in the catalog of \citet{har1996} (2010 revision). For this sample of 59 clusters, we find an average estimated phase space velocity of 120 km s$^{-1}$. Then, considering that the average flight time until `intersection' with the clusters in Figure \ref{fig:infallers} is about 17 Myr, we can say that these clusters will have, on average, moved 2.1 kpc. This is the basis of the error bars in Figure \ref{fig:infallers}.

One object, J024605.10+312254.29, has several orbit realizations consistent with origin in NGC 1904, although it is unlikely that an object with a metallicity of -0.27 dex would originate in a cluster with a metallicity of -1.6 dex. These unbound, infalling, high metallicity objects are some of the most perplexing findings in this study and warrant further investigation.

At this juncture we point out the work of \citet{zie2015} who independently analyzed the \citet{pal2013} sample of objects. After reanalyzing the kinematics, they found that many of the candidates were likely old thick disk objects. The most likely runaway objects were the ones with high ($>$-0.3) metallicity and low $\alpha$ abundances. Thus a logical next step for investigating our candidates would be to obtain the $\alpha$ abundances to try and figure out the chemical characteristics of their origins.

\subsubsection{A Case Against Halo Origin}\label{sec:acaho}

We note that while none of these candidates have more than a 2$\sigma$ chance of being unbound, they all have a greater than 60\% chance of being hypervelocity stars.

We wish to estimate the probability that these objects are merely halo objects which have leaked into our sample. To do this we generate a simulated halo of 10$^{8}$ objects with total space velocities randomly assigned by assuming the same halo velocity parameters used earlier in the paper: $\mu_{R}$ = $\mu_{\Phi}$ = $\mu_{Z}$ = 0 km s$^{-1}$ and $\sigma_{R}$ = 138.2 km s$^{-1}$, $\sigma_{\Phi}$ = 82.4 km s$^{-1}$ and $\sigma_{Z}$ = 89.3 km s$^{-1}$ \citep{smi2009}. This halo velocity profile is then inflated by adding in quadrature the average error on our hypervelocity candidate objects, about 130 km s$^{-1}$.

We take the total space velocities of our candidate objects and assign relative probabilities to them based on where they lie with respect to the distribution of velocities in our toy halo: $P = 1 - b10^{-8}$ where b is the number of halo members with lower speeds than the hypervelocity candidate.

These probabilities range from a few times 10$^{-5}$ to a few times 10$^{-6}$, so, while it is possible that one or even two of these objects could be halo interlopers, it is unlikely that the entire sample is. The ensemble probability of finding seven or more objects at these velocities in our sample is small. If we conservatively estimate that the sample comprises 10\% halo contamination (which is 2.5 times our estimated contamination) the binomial probability of finding seven objects belonging to the halo with these velocities is only 4.6$\times$10$^{-8}$. Even if half of these candidates are spurious, for example from incorrect proper motion measurements, then it is still unlikely that the remaining candidates are halo members - the chance of obtaining three or more such objects from the smooth halo is only 0.4\%.

\begin{figure*}
\includegraphics[width=\textwidth]{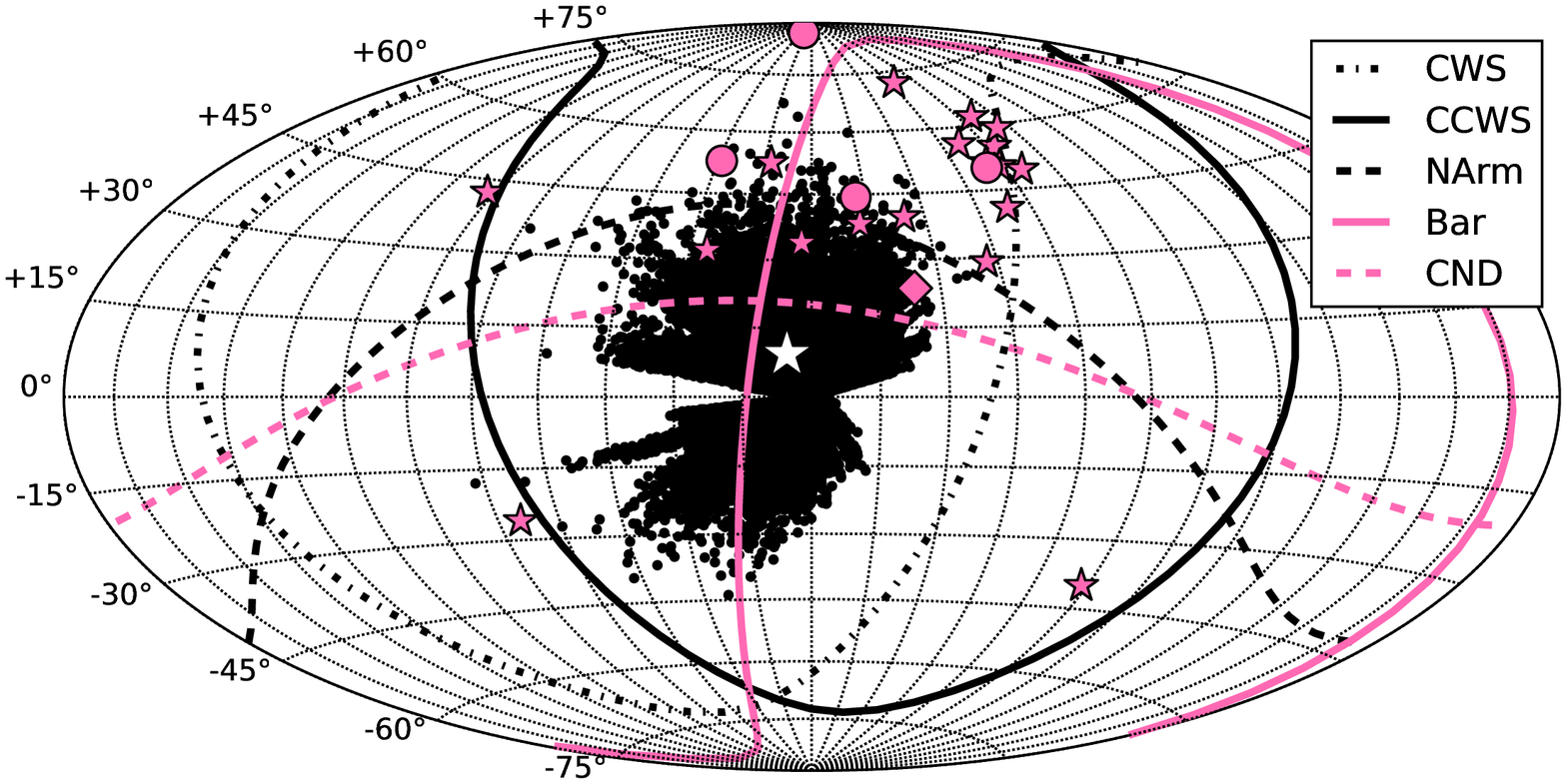}
\caption{Galactocentric Hammer projection of Hills stars with the Sun at 180$^{\circ}$.
The pink stars are Hills stars presented in \citet{bro2012}, the circles indicate the Hills stars from \citet{bro2015} which are distinct from the \citet{bro2012} sample, and the pink diamond is the Hills star recently discovered by \citet{zhe2014}-- these objects are all B type Hills stars, and thus are most likely associated with the star forming planes of matter in the Galactic center.
The white star is Hills Candidate J135855.65+552538.19 (see Table \ref{tab:hills}), the one object in our sample which has an orbit coincident with the Galactic center and velocities which preclude it from being easily explainable with a halo origin.
The lines indicate the orientation of various planes of matter in the Galactic Nucleus as described in \citet{pau2006}. It is expected that young Hills stars will be formed and ejected along these planes while old Hills stars will be ejected in a more isotropic fashion. The black curves indicate planes previously associated with Hills stars. The pink curves are other planes in the Galactic nucleus.}
\label{fig:hills}
\end{figure*}

\subsection{Hills Stars}

\begin{table*}
\begin{center}
\caption{Hills Candidates}
\begin{tabular}{lccccccccc}
Item & IAU Name & $\Delta$ P.M. & Nearest Neighbor & Total P.M. & $\sigma$ P.M. & [Fe/H] & Ejection Velocity & Ejection Radius \\
 &  & mas yr$^{-1}$ & arcseconds &  mas yr$^{-1}$ & mas yr$^{-1}$ & dex & km s$^{-1}$ & kpc \\
\tableline
1 & J131247.03-011414.01 & 4.2 & 14.36 & 31.81 & 4.21 & -0.67 & 455.19$\pm$91.77 & 0.42$\pm$0.84 \\
2 & J135855.65+552538.19 & 4.42 & 13.33 & 60.24 & 3.78 & -0.78 & 637.28$\pm$62.85 & 0.2$\pm$0.63 \\
3 & J142648.59+563316.32 & 8.85 & 13.98 & 19.82 & 3.9 & -0.61 & 476.18$\pm$98.74 & 0.15$\pm$0.99 \\
4 & J165150.71+242750.95 & 3.26 & 7.33 & 56.55 & 3.76 & -0.64 & 475.19$\pm$67.51 & 0.37$\pm$0.46 \\
5 & J160403.12+332931.07 & 0.94 & 10.66 & 10.12 & 3.59 & -0.71 & 453.19$\pm$105.56 & 0.45$\pm$0.97 \\
6 & J233657.12-002138.79 & 5.38 & 18.63 & 13.32 & 3.62 & -0.77 & 463.91$\pm$123.09 & 0.08$\pm$1.0 \\
7 & J182730.89+211433.40 & 2.66 & 9.32 & 3.09 & 4.11 & 0.08 & 461.2$\pm$91.22 & 0.24$\pm$0.82 \\
8 & J171543.37+431736.88 & 12.52 & 4.17 & 14.27 & 3.73 & -0.6 & 472.37$\pm$124.58 & 0.29$\pm$0.89 \\
9 & J113946.33-033651.63 & 1.61 & 12.62 & 68.38 & 3.62 & -0.5 & 481.68$\pm$100.01 & 0.11$\pm$0.64 \\
10 & J000650.47-071035.21 & 4.22 & 7.09 & 26.78 & 3.58 & -0.64 & 492.21$\pm$83.14 & 0.19$\pm$0.64 \\
11 & J000613.77-051228.28 & 4.02 & 6.76 & 19.48 & 3.28 & -0.77 & 459.76$\pm$106.11 & 0.3$\pm$0.9 \\
\tableline
\label{tab:hills}
\end{tabular}
\tablecomments{Our collection of objects leaving the central half kpc of the Galaxy (radius) at more than 450 km s$^{-1}$ with sub-kpc crossing errors. The columns $\Delta$ P.M. (the difference in the SDSS and PPMXL proper motion estimates), distance to the nearest neighbor (in projection), total proper motion and proper motion error are provided to gauge the reliability of the measurements (see Appendix A). The last two columns are the velocity of ejection and the distance from the Galactic center.}
\end{center}
\end{table*}

Finally we investigate objects which may have been ejected by interactions with the central supermassive black hole \citep{hil1988}. There are many exciting questions that can be answered by investigating possible Hills stars.

Some authors postulate that the central supermassive black hole may in fact be a binary system. \citet{lu2007} claim that an ejected, undisrupted stellar binary system would be a `smoking gun' indication that there are two black holes orbiting each other in the center of our Galaxy and would constitute one of the best indications of such a phenomenon without gravitational wave experiments. However, \citet{per2009} notes that such findings are probably only true in the low mass regime simulated by \citet{lu2007} and investigated here. In the high mass realm, ejected binary systems could be much more common owing to the differing frequencies of triple systems in low and high mass stars. In fact, ejected binary systems provide a convenient explanation for so called `too young' runaways -- stars whose main sequence lifetimes are shorter than their flight times -- through mass feeding rejuvenation of the lower mass companion. Still, a low mass hypervelocity binary could be indicative of a binary black hole system.

Another is the interesting discovery that hypervelocity stars have a non-isotropic distribution \citep{bro2012}. It has been suggested by \citet{lu2010} that the locations of these ejected stars are related to the planar structures that they existed in before ejection and, as a consequence, they should not be distributed isotropically.

In the Galactic center, it has been shown by \citet{pau2006} that some disks of material falling to the supermassive black hole host star formation (specifically a ClockWise System and a CounterClockWise System [CWS, CCWS]) and that these disks are rich in a top-heavy stellar population. Stars from these systems are ejected preferentially in paths parallel with their birth planes. Other systems, such as the Northern Arm and Bar of the minispiral and the CircumNuclear Disk (NArm, Bar, CND) also exist and star formation in these features could also eject young stars in planar distributions.

Previously studied B type Hills stars, which are necessarily young objects, are most likely to be found in planar orientations since they were probably born in these infalling structures shortly before ejection. Older populations in the Galactic center probably fall into the nuclear region along random vectors owing to the chaotic nature of the surrounding bulge. And some stars born in the above mentioned structures may also be scattered into random orbits if they survive long enough. This is observationally confirmed by \citet{fig2003}.

The general picture of the nuclear regions is then a highly ordered and coherently rotating young stellar population and an isotropically distributed old population. If we see objects being ejected from the Galactic center, we would expect them to be a mixture of planar orientation and random distribution since our F-to-M type stars can cover a wide spread of ages.

In Table \ref{tab:hills} we present a collection of candidate objects which left the central regions of the Galaxy at high velocities. Initially we construct a sample of objects whose nominal orbits had them crossing the central 1.5 kpc (radius) of the Galaxy at more than 350 km s$^{-1}$ -- 452 candidates. These objects are then integrated through the Galactic potential back in time for 50 orbit realizations to calculate the errors on their crossing positions and crossing velocities. In general, the ensemble has a lower average crossing velocity than the nominal orbits. When we reduce the sample to a subset of objects calculated to have average crossing velocities greater than 450 km s$^{-1}$, which cross the central half kpc, and also have sub-kpc crossing position errors, we retain a sample of just 11 high quality Hills object candidates.

Here we wish to estimate the probability that these objects are merely extremely hot bulge stars in our sample. We implement a methodology similar to that presented in Section \ref{sec:acaho}. The approach is the same except for the dispersion of our toy model. The toy bulge is defined to have a velocity dispersion of 102.2 km s$^{-1}$: this number being drawn from the relationship between black hole mass (about 4.3 10$^{6}$ M$_{\odot}$ as calculated by \citealt{gil2009}) and central velocity dispersion as calculated for spiral galaxies by \citet{mcc2011}. This value is in agreement with the findings of \citet{nes2013} for the velocity dispersion of metal poor stars in the Milky Way bulge and is on the high end for metal rich stars. The individual probabilities of these objects belonging to the bulge are all on the order of 10$^{-5}$ except for object 3, J135855.65+552538.19, which has a probability of 10$^{-8}$. If we conservatively assume that the bulge forms 5\% of our total sample, we estimate that the ensemble probability of finding six or more bulge stars with velocities consistant with our Hills candidates is extremely low (at around 4$\times$10$^{-6}$).

While it is unlikely that these objects come from the bulge, it is possible that they are halo stars. To test this, we first check the radial domination of their orbits via the $\beta$ parameter:

\begin{equation}
\beta = 1 - \frac{v_{\theta}^2 + v_{\phi}^2}{2v_{R}^2},
\label{eqn:beta}
\end{equation}

compared to that of a simulated halo (with a velocity profile as that in \citealt{smi2009}). We find that, while these objects in general have more radial orbits than our simulated halo, the evidence is not overwhelming. The objects are generally more radial than 75\% of our simulated orbits with the exception of J135855.65+552538.19, whose orbit is more radial than 99\% of our simulated halo. As another test, we generate a 13,500 halo objects (a generous 10\% of our sample) in our observational area and integrate their orbits back in time to estimate how many have crossing velocities and positions similar to our Hills candidates. This test also implies that our Hills candidates could be halo stars, with the exception again of J135855.65+552538.19. In our simulated halo, no stars cross with a velocity greater than 630 km s$^{-1}$ and a radius $<$ 0.25 kpc.

In Figure \ref{fig:hills}, we compare the object's location (white star) with the planes presented in \citet{pau2006}. This plot is a Galactocentric longitude and latitude plot with the sun located at 180$^{\circ}$ in longitude: this choice of coordinate system makes spatial orientation correlations with Galactocentric features more obvious than they would be in a heliocentric system. In this figure, we also plot the spatial locations of the hypervelocity stars presented in \citet{bro2012} as well as the recently discovered hypervelocity star of \citet{zhe2014}. As has been noticed before, the B type hypervelocity stars in general lie near to the planes occupied by the CWS, NArm and CCWS. Our survey footprint is shown in black. We find that the location of our Hills candidate is close, but not exactly coincidental with the plane of the bar of the minispiral structure in the nuclear regions, thus its origin is ambiguous and it needs further study. As far as we can tell though, this object is not easily explainable with halo, disk, or bulge origin and constitutes our best candidate for Hills mechanism origin. Unfortunately, since we have a sample of one, we cannot comment if this object was ejected from an ordered plane like accretion or an isotropic cloud like environment.

It is interesting to note that \citet{kol2010} place upper limits on the ejection rate of solar metallicity F/G type dwarfs at no more than 30 times that of the young B type ejecta. Despite being an upper bound, the total number of known B type Hills ejecta is $\sim$21 \citep{bro2014}; so even considering completeness limits of the SDSS spectroscopic sample utilized in this study, it is not completely unfeasible that a true F-M type Hills ejectum has been observed.

Follow up observations of this object, combined with existing observations of the known Hills stars, could lead to valuable insight into the shrouded populations of the innermost regions of the Galaxy. Spectroscopic observations of the stars considered in tandem with existing samples of B type Hills stars could also potentially provide insights into the chemical abundances of the various nuclear structures. Furthermore, as mentioned above, if we could discover an ejected, coherent binary system this would provide groundbreaking evidence for the existence of a binary black hole at the center of the Milky Way \citep{lu2007}.

It is worth noting that two objects in Table \ref{tab:hills} have coincident positions with a 65 pc separation, but their orbits are not aligned with each other.

\section{Conclusion}

We have collected a sample of high metallicity ($>$ -0.8 dex) main sequence F-to-M type stars. Using six dimensional phase space information for these objects we constructed probability density functions describing the bulk motion as a function of height from the Galactic plane. These probability density functions were then used to identify kinematic outliers.

We found that outliers are generally faster moving, more metal-poor and of earlier stellar type than natural stars. Outliers also tend to show a preference for originating closer to the Galactic center than the natural stars. This implies that the outlier stars are poorly described by a halo population, since halo contamination would be spatially isotropic. Of our 5\% least likely objects, for example, 54\% have positive radial velocities with respect to the Galactic center (and 58\% of our 1\% least likely). Our metallicity cut should also minimize the contribution of halo stars to our sample. This is consistent with findings of \citet{nes2013}, who find that the bulge component with a metallicity of -1.0 $<$ Fe/H $<$ -0.5 is kinematically hotter than the higher metallicity components.

It is interesting that the earlier type stars are kinematically hotter than the later type stars. One explanation for these trends could be that the binary fraction is higher for earlier type stars. \citet{kou2009} notes from a literature compilation that F to G type dwarfs have multiplicities of 55\%-60\% while M dwarfs have multiplicities of 30\%-40\% and late M types and brown dwarf binary fractions can be as low as 10\%-30\%. This higher binary fraction enables more dynamic ejections, as the ejected star usually originates in a binary system. A higher existent binary fraction could also explain the generally more dispersed velocity components seen in Figure \ref{fig:f_vs_k}, as the binary orbit velocity components would add to the system's velocity components.

We have also collected a sample of hypervelocity stars calculated to be traveling faster than their local escape velocity. After imposing stringent cuts designed to reduce proper motion measurement errors, we integrate their orbits back in time to find their origins. Of our 7 strong candidates, two of four are found to intersect the Galactic plane at positions coincident with spiral arm structure (one intersects relatively distantly to be consistent with disk origin and between spiral arms and one crosses the Galactic plane too distantly to be associated with any known structure). It is well known that runaway star production is higher in the spiral arms \citep{sil2013} due to the greater stellar density. The remaining three hypervelocity stars have orbits that indicate the objects are actually infalling, rather than being ejected from the disk. We perform Monte Carlo error expansions on their orbits to try and uncover possible origins: one object has an orbit which intersects NGC 1904, but the metallicity difference between the star and the globular cluster pose a mystery. Such infalling hypervelocity stars have been noticed recently by \citet{pal2013} as well.

Finally we examined a sample of stars leaving the central half kpc at 450 km s$^{-1}$ or greater. These stars may have interacted with the central supermassive black hole and received large kicks from that interaction \citep{hil1988}. Hills stars may come to encounter the central supermassive black hole via two paths: younger stars may be formed on gaseous disks in the nuclear region; and older stars may fall in from outside the nuclear region isotropically. Detailed tests of their orbits compared to those of simulated halo objects cannot rule out halo origins for all of these stars with the exception of one strong candidate: J135855.65+552538.19. This object has an extremely fast, radial orbit, intersects the nuclear regions, and lies close to one of the expected nuclear planes; however, we cannot comment on its origin being from this plane without a larger sample size and more precise observations.

\section{Acknowledgments}

We thank Drs Avon Huxor, Stephen Justham, Youjun Lu, Anna Pasquali, Siegfried R\"{o}ser, Qingjuan Yu, and Zheng Zheng for useful discussion and helpful explanations. We thank the developers and maintainers of the following software libraries which were used in this work: NumPy, SciPy, astroML \citep{van2012}, PyMC, matplotlib and Python.

This work was supported by the Marie Curie Initial Training Networks grant number PITN-GA-2010-264895 ITN ``Gaia Research for European Astronomy Training" and by Sonderforschungsbereich SFB 881 ``The Milky Way System" (subproject A2 and A3) of the German Research Foundation (DFG). M.C.S. acknowledges financial support from the CAS One Hundred Talent Fund and from NSFC grants 11173002 and 11333003. This work was also supported by the Strategic Priority Research Program The Emergence of Cosmological Structures of the Chinese Academy of Sciences, Grant No. XDB09000000 and the National Key Basic Research Program of China 2014CB845700. J.J.V. is a fellow at the International Max Planck Research School for Astronomy and Cosmic Physics at the University of Heidelberg and a member of the Heidelberg Graduate School for Fundamental Physics; JJV also acknowledges the support of a LAMOST fellowship.

The SDSS-III Collaboration (www.sdss3.org) includes many institutions from around the globe. Funding for SDSS-III has been provided by the Alfred P. Sloan Foundation, the Participating Institutions, the National Science Foundation, and the U.S. Department of Energy. The SDSS-III is managed by the Astrophysical Research Consortium for the Participating Institutions of the SDSS-III Collaboration including the University of Arizona, the Brazilian Participation Group, Brookhaven National Laboratory, University of Cambridge, University of Florida, the French Participation Group, the German Participation Group, the Instituto de Astrofisica de Canarias, the Michigan State/Notre Dame/JINA Participation Group, Johns Hopkins University, Lawrence Berkeley National Laboratory, Max Planck Institute for Astrophysics, New Mexico State University, New York University, the Ohio State University, the Penn State University, University of Portsmouth, Princeton University,the Spanish Participation Group, University of Tokyo, the University of Utah,Vanderbilt University, University of Virginia, University of Washington, and YaleUniversity.

The results presented in this publication make use of data from the Two Micron All Sky Survey (2MASS), which is a joint project of the University of Massachusetts and the Infrared Processing and Analysis Center, funded by the National Aeronautics and Space Administration and the National Science Foundation.

\clearpage

\begin{appendix}

\section{Appendix A: Assessment of Proper Motion Accuracy}

Errors in the proper motion measurements are the most likely cause of misidentified hypervelocity stars and Hills ejecta. While the typical SDSS radial velocity errors are less than 15 km s$^{-1}$ for all of our objects \citep{yan2009}, a proper motion error of 3 mas yr$^{-1}$ is about 15 km s$^{-1}$ kpc$^{-1}$. Therefore it is crucial that we analyze the reliability of the SDSS proper motion measurements and the SDSS proper motion error estimates.

To investigate this, we utilize a variety of data sources. For every object in the text (specifically in Sections 4.2 and 4.3), we have four values which can be obtained: their PPMXL proper motions (PPMXL is a 2 Micron All Sky Survey based proper motion catalog; \citealt{skr2006}, \citealt{roe2010}), their SDSS proper motions \citep{mun2004}, their SDSS proper motion error estimates, and the distance to their nearest neighbor on the sky.

For a small part of the sky, SDSS stripe 82\footnote{Stripe 82 is a section of sky along the southern Galactic cap that was reobserved in 303 runs for the purpose of transient object investigation; see http://www.sdss.org/legacy/stripe82.html}, we have two additional proper motion catalogs based solely on SDSS photometry: that of \citet{bra2008} and that of \citet{kop2013}.
We attempt to use those four `quality metrics' which we have for the entire study:
\begin{itemize}
\setlength{\itemsep}{1pt}
\item The total difference between the SDSS proper motions of \citet{mun2004} and the PPMXL proper motions of \citet{roe2010}: $\Delta$ (Munn, Roeser),
\item The total pipeline proper motions of the SDSS: $|$P.M. (Munn)$|$,
\item The distance on the sky to the nearest SDSS detection: Nearest Neighbor ("),
\item The pipeline reported error on the proper motions of the \citet{mun2004} catalog: $|$P.M. Error (Munn)$|$,
\end{itemize}
to create a reliability metric by comparing the four proper motion catalogs (\citealt{mun2004}, \citealt{roe2010}, \citealt{kop2013}, \citealt{bra2008}) in the common area of SDSS stripe 82.

\begin{figure*}[h!]
\includegraphics[width=\textwidth]{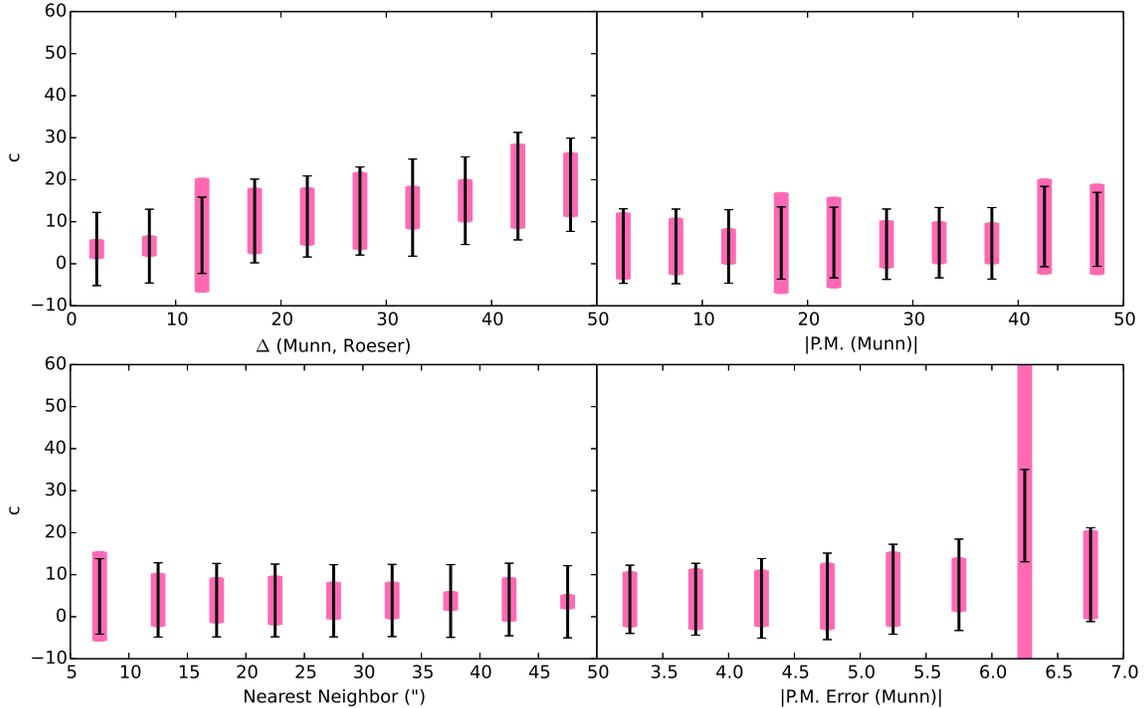}
\caption{Average clustering as a function of the four `quality metrics.' In pink we show the one sigma dispersion in c (the `clustering' of proper motion measurements in different surveys, described in the text; small c indicates good agreement between various proper motion studies). In black, we show additional error bars centered on the average value of the clustering, c. These error bars' extents are the average value of the catalogs' errors (added in quadrature) for each bin. We see that the clustering grows gradually worse as the disagreement between the catalogs of \citet{mun2004} and \citet{roe2010} increases. No significant trend is apparent as a function of total proper motion, although larger reported errors do indicate larger spreads in the clustering. As expected, there is better agreement for all of the proper motion catalogs in less crowded areas.}
\label{fig:pm_1}
\end{figure*}

Our first metric of accuracy is the `clustering' of the four measurements:
$$c = \sqrt{\sigma(P.M._{R.A.})^{2} + \sigma(P.M._{Dec.})^{2}. }$$

In Figure \ref{fig:pm_1} we plot the clustering c as a function of the four `quality metrics'. We find that the four proper motions are in best agreement when the PPMXL and SDSS proper motions are similar, when the reported errors are small, and the field is relatively uncrowded.

\begin{figure*}[h!]
\includegraphics[width=\textwidth]{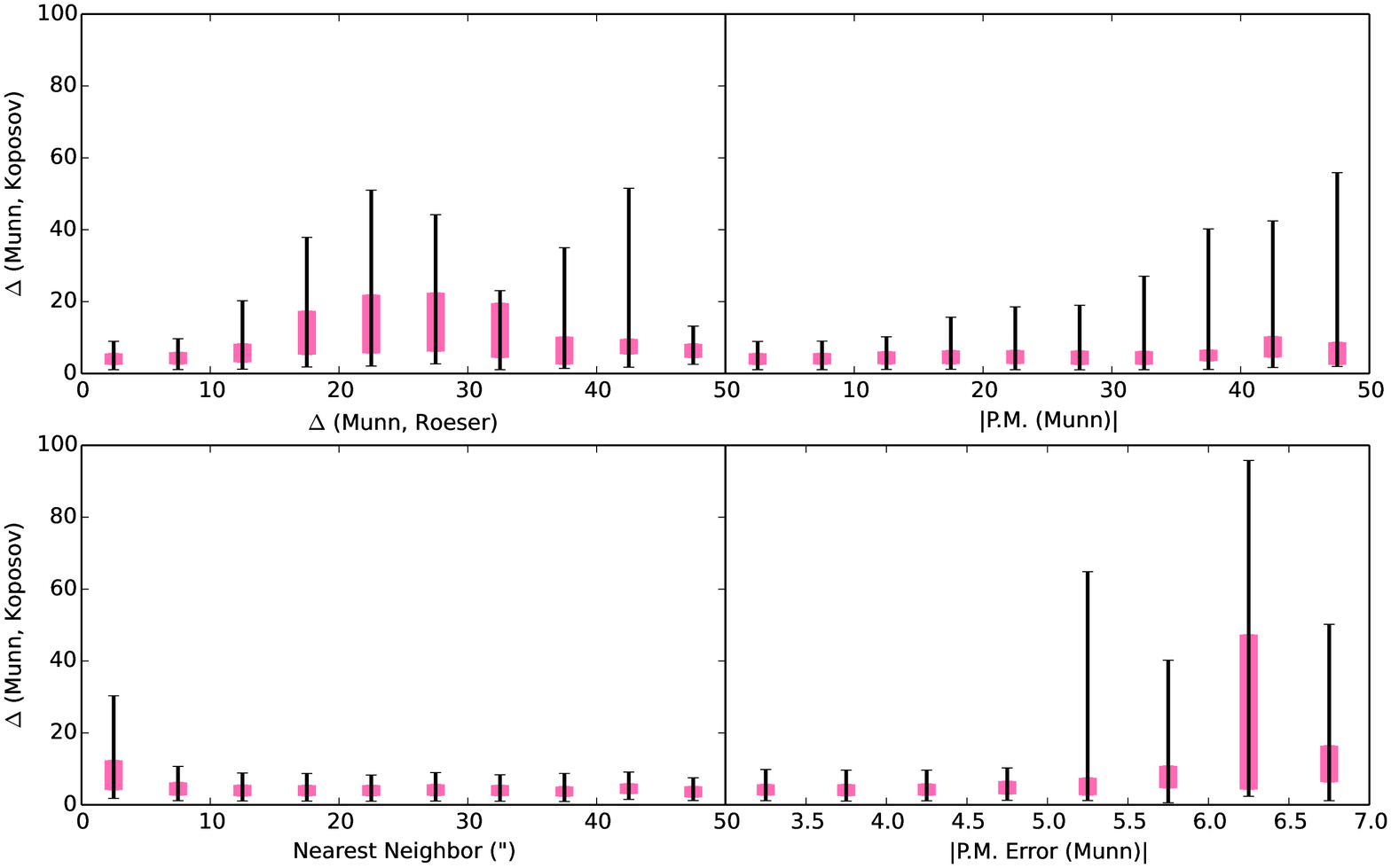}
\caption{Difference in proper motions for objects in the proper motion catalogs of \citet{kop2013} and \citet{mun2004}. The pink and black bars indicate the interquartile and 95\% confidence ranges, respectively. Visible in this figure is a definite trend for agreement to become more erratic as a function of \citet{roe2010} and \citet{mun2004} catalog disagreement and the total proper motion reported by \citet{mun2004}. There is also good agreement for measurements in uncrowded fields and in measurements with low reported errors.}
\label{fig:pm_2}
\end{figure*}

As another test, in Figure \ref{fig:pm_2} we check the difference between the highly precise catalog of \citet{kop2013} and the SDSS pipeline catalog. The catalog of \citet{kop2013} is a wholly internal catalog, using only SDSS photometry, so it avoids systematic mismatching biases caused by crossmatching catalogs. Many proper motion catalogs, such as those of \citet{mun2004} and \citet{roe2010}, are matched to photographic plate surveys to achieve long time baselines, at the cost of possible source mismatching. This two catalog check will lessen the possibility of `clustering inflation' caused by intrinsic catalog mismatches in the prior test. Again we see that large proper motion errors indicate noisier agreement, and that the best agreement is in uncrowded fields.

In light of Figure \ref{fig:pm_1} and Figure \ref{fig:pm_2}, we suggest the constraints presented in Table \ref{tab:pm} for the selection of high quality proper motion measurements. We use these cuts to select candidate hypervelocity stars in Section 4.2.

\begin{table*}[h0]
\begin{center}
\caption{}
\begin{tabular}{ccc}
\tableline\tableline
Constraint & $\Delta$ (Munn, Koposov) & c \\
 & mas yr$^{-1}$ & mas yr$^{-1}$\\
 \tableline
$|$SDSS P.M. - PPMXL P.M.$|$ $<$ 12  & 4.5  & 4.0  \\
$|$SDSS P.M. - PPMXL P.M.$|$ $>$ 12  & 14.6  & 21.5  \\
\tableline
$|$ SDSS P.M. $|$ $<$ 30  & 4.7  & 4.5  \\
$|$ SDSS P.M. $|$ $>$ 30  & 18.8  & 11.7  \\
\tableline
Nearest Neighbor $>$ 10" & 4.3  & 4.1  \\
Nearest Neighbor $<$ 10" & 6.6  & 5.9  \\
\tableline
SDSS P.M Error $<$ 5  & 4.7  & 4.5  \\
SDSS P.M Error $>$ 5  & 16.4  & 13.1  \\
\tableline
\label{tab:pm}
\end{tabular}
\tablecomments{The average difference between the proper motion estimates of \citet{kop2013} and the SDSS pipeline for cuts on certain constraints. The average clustering, c (described in the text), is also shown. The cuts here are chosen by inspection of Figures \ref{fig:pm_1} and \ref{fig:pm_2}.}
\end{center}
\end{table*}

\begin{figure*}
\includegraphics[width=\textwidth]{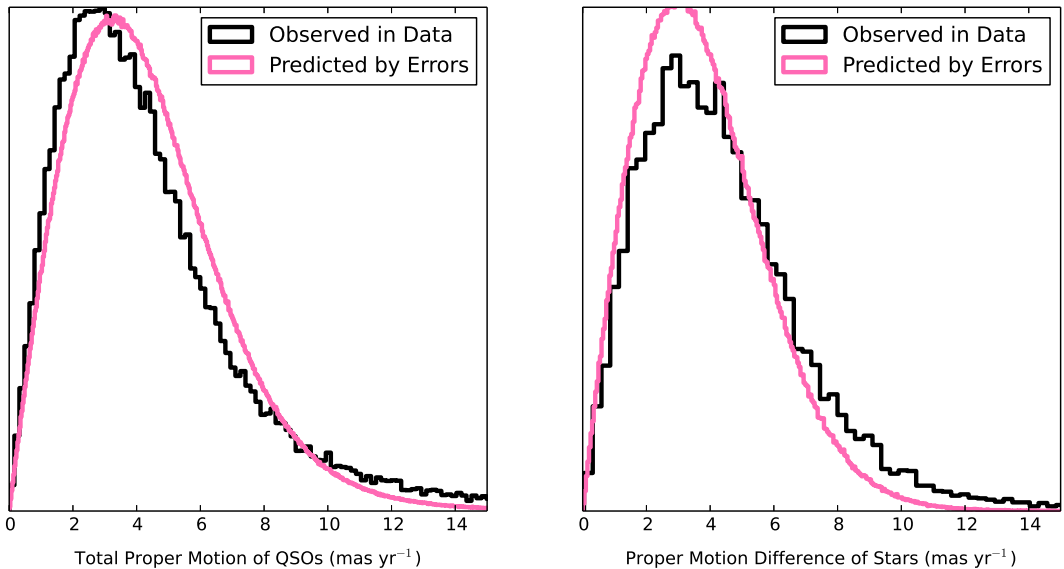}
\caption{Left: comparison of the observed proper motions of quasistellar objects (QSOs) in the proper motion catalog of \citet{mun2004} (black) and the proper motions  predicted by errors (pink). Since QSOs should be essentially stationary on the sky, their measured motion is solely due to observational uncertainties. We note that the predicted motion from errors is in general larger than the observed motion, which implies that the errors are being accurately reported or even overestimated. \\
Right: A comparison of the difference of proper motions in the \citet{kop2013} and \citet{mun2004} catalogs (black) along the the proper motion differences predicted by the errors of the two catalogs (pink). In general there is good agreement; there is a slightly heavy tail in the observed differences, this is most likely an effect of mismatching between the two catalogs.}
\label{fig:pm_errs}
\end{figure*}

Having investigated the accuracy of the proper motions, we now investigate the accuracy of the proper motion errors. This is a critical aspect as it heavily affects the outcomes of our Monte Carlo expansions. To do this we again compare the catalogs of \citet{kop2013} and \citet{mun2004}. In Figure \ref{fig:pm_errs} we plot the observed difference in stellar proper motions along with the expected differences from the reported errors of both catalogs. As a secondary test, also in Figure \ref{fig:pm_errs}, we plot the motions of spectroscopically identified quasars in the catalog of \citet{mun2004} along with their expected motions as a result of the errors on their proper motion estimates.

For the quasars, we see that the proper motions predicted by the formal proper motion uncertainties are marginally larger than the observed ones, implying that Munn's errors may be slightly overestimated. In the comparison of proper motions between the two catalogs we see, in general, a reasonably good agreement between the observed and predicted distributions. The observed difference is slightly larger than the formal uncertainties would suggest, but this is not a strong effect. The small tail in this plot is likely an effect of mismatching the objects between the catalogs and, in conclusion, we believe that the errors reported in the \citet{mun2004} catalog are reliable.

\end{appendix}

\end{document}